\documentclass[twocolumn,showpacs,preprintnumbers,amssymb,pre,superscriptaddress]{revtex4}

\usepackage{graphicx}
\usepackage{dcolumn}
\usepackage{bm}
\usepackage{amstext}

\begin{document}


\title{The role of body rotation in bacterial flagellar bundling}

\author{Thomas R. Powers}
\email{Thomas_Powers@brown.edu}
\affiliation{%
Division of Engineering, Box D, Brown University, Providence, RI
02912
}%

\date{January 28, 2002}


\begin{abstract}
In bacterial chemotaxis, {\it E. coli} cells drift up chemical
gradients by a series of runs and tumbles. Runs are periods of
directed swimming, and tumbles are abrupt changes in swimming
direction. Near the beginning of each run,
the rotating helical flagellar filaments which propel the
cell form a bundle. Using resistive-force theory, we show that the
counter-rotation of the cell body necessary for torque balance is
sufficient to wrap the filaments into a bundle, even in the
absence of the swirling flows produced by each individual
filament.
\end{abstract}


\maketitle

 Although bacteria are among the simplest systems for
the study of cell motility, many puzzles remain.  Chief among
these is the mechanics of the bundling and unbundling of flagellar
filaments in the chemotaxis behavior of bacteria such as
\textit{E. coli} and
\textit{Salmonella}~\cite{macnab_ornston1977,bray1992}.  These
cells move toward higher concentrations of favorable chemicals by
executing a series of runs and tumbles~\cite{berg1993}. The runs
are periods of directed swimming. At the end of each run, the cell
randomizes its direction by tumbling.  If the cell happens to head
in a favorable direction, the likelihood of tumbling reduces,
making runs in this direction longer on average compared to runs
in the unfavorable direction. Propulsion during a run is generated
by the rotation of several helical propellers, known as flagellar
filaments. Unlike eukaryotic flagella~\cite{bray1992}, bacterial
flagellar filaments are passive elements driven by rotary motors
embedded in the cell wall. Near the beginning of a run, the motors
turn in a counter-clockwise direction (when viewed from the
outside of the cell), and the left-handed filaments come together
to form a bundle. At the end of a run, one or more of the motors
reverses, and one or more of the filaments fly out of the bundle
and cause the cell to tumble. This process is complex and involves
changes in filament handedness and pitch.  The cell soon sets out
on a new course but regains its initial speed only after the
aberrant motors have reversed again and their filaments have
regained their normal conformation and rejoined the
bundle~\cite{turner_ryu_berg2000}.

Although qualitative partial explanations for bundle formation
have appeared in the
literature~\cite{anderson1975,macnab1977,lighthill1975}, a
mathematical theory has not.  In this note we begin to construct
this theory with a quantitative treatment of one aspect of the
bundling phenomenon: the role of the cell body rotation. This
rotation and the accompanying hydrodynamic resistance arise to
balance the torque exerted by the rotating bundle on the cell
body. Thus, in the body-fixed frame, there are two kinds of flows
which contribute to bundling: the flow due to frame rotation, and
the swirling flows set up by each individual filament.  Here we
focus on the flow due to rotation of the body-fixed frame;
the swirling flows and
interactions among flagellar filaments will be treated in a
separate publication~\cite{promise}.  Our treatment is in the
spirit of Machin~\cite{machin1958}, who used a similar approach to
argue that eukaryotic flagella could not be passive elements
driven by motors at the cell body (see also
ref.~\cite{wiggins_goldstein1998}).

\begin{figure}
\includegraphics[height=1.8in]{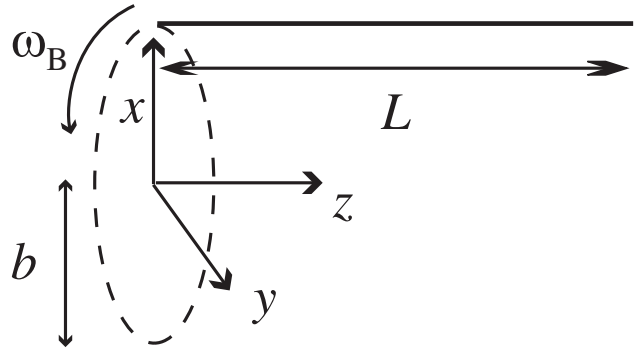}
\caption{\label{setupfig} Model problem.  A naturally straight but
flexible rod, initially parallel to the $z$-axis, with one end
held fixed with zero moment at $x=b$, $y=0$, $z=0$. We seek the
steady-state shape when the end of the rod is forced to rotate in
the $x$-$y$ plane about the $z$-axis at frequency
$\omega_{\textrm{B}}$.}
\end{figure}

Fig.~\ref{setupfig} illustrates the model problem.
For simplicity, replace the helices with straight but flexible
rods of length $L$, rotated with the frequency $\omega_{\textrm{B}}$
about the cell body axis of symmetry, $z$. Let $b$ denote the
distance between the axis of the unstressed rod and the $z$-axis.
Since the body is about a micron across, and flagellar filaments
are typically six to ten microns long, we suppose $b\ll L$.  We
also disregard the rotational \textit{disturbance} flow arising from
the no-slip condition at the cell body.   In the body-fixed frame,
this disturbance flow reduces the net rotational flow near the body.
We disregard this disturbance flow 
since the flow field of a sphere of radius $a$ rotating
with angular velocity $\bm{\omega}$ takes the form
$\bm{v}=\bm{\omega}\times\bm{r}(a/r)^3$, falling off rapidly with
$r$~\cite{happel_brenner1965}.  Likewise, it is argued below that
the axial drag on a filament due to the nonzero swimming velocity
plays little role in our problem. Finally, we focus our attention
on the contribution to flagellar filament wrapping due to body
rotation, and not the flows set up by the individual rotating
filaments, by ignoring the hydrodynamic interactions among the
rods. Thus, it suffices to consider the shape of a single rod.

\begin{figure}
\includegraphics[height=3.0in]{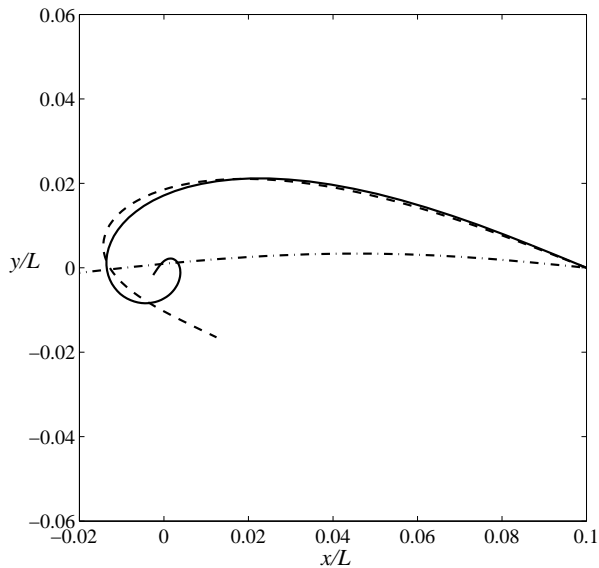}
\caption{\label{shapesxy} Projection of shapes of a rotating
flexible rod onto the $x$-$y$ plane for $\ell/L=0.1$ (solid line),
$\ell/L=0.2$ (dashed line), and $\ell/L=0.5$ (dot-dashed line).}
\end{figure}

During runs, the left-handed flagellar filaments turn
counter-clockwise (when viewed from outside the cell), and the
body turns clockwise (when viewed from behind---i.e. from the
distal end of the bundle). When our model filament is turned about
the body-rotation axis $z$ in this same sense (clockwise when
viewed from the positive $z$-axis, see fig.~\ref{setupfig}), it
forms a right-handed shape (e.g. see fig.~\ref{shapesxy} and note
that the proximal end $x/L=b=0.1$ is in the plane $z=0$, and the
distal end with $x$
near $0$ has positive $z$ coordinate). 
Furthermore, two lefthanded helices rotating about their
respective axes with proximal ends held stationary will lead to a
flow which also tends to wrap the helices around each other in a
righthanded manner~\cite{promise}. Thus, the body-rotation effect
treated here and the swirling flow effect treated
in~\cite{promise} act to wrap the filaments in the same sense.

Since typical Reynolds numbers for swimming bacteria are of order
$10^{-6}$~\cite{berg1993}, inertia is unimportant and the
steady-state rod shape is determined by a balance of viscous and
elastic forces per unit length. For gentle distortions of a slender 
body, the
viscous forces per unit length are well-approximated by the 
resistive-force coefficients (\textit{e.g} see~\cite{lighthill1975}
and references therein):
$\bm{f}=\zeta_\perp\bm{u}_\perp+ \zeta_{||}\bm{u}_{||},$ where
$\bm{u}_\perp$ and $\bm{u}_{||}$ are the perpendicular and
parallel components of the local rod velocity relative to the
fluid velocity $\bm{v}$: $\bm{u}=
\partial\bm{r}/\partial t-\bm{v}$. The transverse friction
coefficient (per unit length) is of the form
\begin{equation}
\zeta_\perp\approx{4\pi\eta\over\log(L/a)+1/2} , \label{zetaperp}
\end{equation}
where $\eta$ is the fluid viscosity, $a$ is the rod radius, and
$\log$ denotes the natural logarithm~\cite{lighthill1975}. As
discussed below, $\zeta_{||}$ does not enter the analysis since we
work in the linearized approximation. Resistive-force theory gives an
accurate value for the drag force per unit length on a slender filament
except near the filament ends; however, the effect of this error on the
shape is ${\cal O}(a/L)$~\cite{higdon}. There is
also a viscous torque tending to twist the rod; however, the
effects of this torque are subleading compared to the effects of
the translational drag ~\cite{wolgemuth_et_al2000a}. To see why,
note that the total torque from rotational drag is ${\cal
O}(\omega\zeta_rL)$, where $\zeta_r=4\pi\eta a^2$ is the friction
coefficient for rotation~\cite{landau_lifshitzFM}. The total
torque from translational drag is ${\cal O}(b(\omega
b\zeta_\perp)\ell)$, where, as we shall see below, only the
portion of the rod within a distance $\ell$ of the held end
$(z=0)$ contributes to the translational drag.  The ratio of these
two torques is ${\cal O}((a/b)^2(L/\ell)[\log(L/a)+1/2])$. For the
representative values $L=10$ $\mu$m, $a \approx10$ nm, and
$b\approx1$ $\mu$m, this ratio is small even if $L/\ell\approx
10$. We therefore disregard rotational drag and twist strain.

To find the bending force per unit length, note that since $b\ll
L$, the displacement of any rod element will also be small.  Thus,
the elastic energy is well-approximated by the quadratic
expression
\begin{equation}
{\cal E} ={1\over2}A\int\bigg[\bigg({\partial^2x\over\partial
z^2}\bigg)^2+\bigg({\partial^2y\over\partial z^2}\bigg)^2\bigg]\
\textrm{d}z, \label{elastic_energy}
\end{equation}
where $x$ and $y$ are as in fig.~\ref{setupfig}, and $A$ is the
bend modulus~\cite{landau_lifshitz_elas}.  Since the variation in
rod shape is rapid for sufficiently high rotation rate, even for
small $b$, this approximation eventually fails and must be
replaced by the full geometrically nonlinear elastic rod energy.
As we discuss below, the rotation rates of interest are small
enough for~(\ref{elastic_energy}) to hold. The variational
derivative of~(\ref{elastic_energy}) yields the elastic bending
force per unit length: $-\delta{\cal E}/\delta\bm{r}_\perp
=-A\partial^4\bm{r}_\perp/\partial z^4,$ where $\bm{r}_\perp\equiv
x \hat{\bm{x}}+y\hat{\bm{y}}$. 

To leading order for $b\ll L$, the motion of the rod is purely
perpendicular to the rod centerline, yielding the equation of
motion~\cite{machin1958}
\begin{equation}
\zeta_\perp\bigg({\partial\bm{r}_\perp\over\partial t}
-\bm{v}_\perp \bigg)=-A{\partial^4 \bm{r}_\perp\over\partial z^4},
\label{eom}
\end{equation}
where $\bm{v}_\perp$ is the transverse fluid velocity. Since
inertia is unimportant in the limit of zero Reynolds number,
eqn.~(\ref{eom}) applies equally well to the rotating frame in
which the rod is fixed and the flow is
$\bm{v}_{\perp}=\omega_{\textrm{B}}\hat{\bm{z}}\times\bm{r}_{\perp}$.
Such a flow tends to wrap the rod around the $z$-axis in a shape
with a helical modulation and exponential envelope. In the
steady-state, eqn.~(\ref{eom}) reduces to
\begin{eqnarray}
-y&=&\ell^4{\partial^4 x\over\partial z^4}\label{carteom1}\\
x&=&\ell^4{\partial^4 y\over\partial z^4}, \label{carteom2}
\end{eqnarray}
where $\ell\equiv(A/\zeta_{\perp}\omega_{\textrm{B}})^{1/4}$ is the
characteristic length scale associated with bending and
drag~\cite{machin1958}. The solution to
eqns.~(\ref{carteom1},\ref{carteom2}) is a simple generalization
of Machin's solution to the in-plane bending problem:
\begin{equation}
x(z)=\sum_{n=1}^8 A_n \exp(r_n z/\ell), \label{machin_solns}
\end{equation} where $r_1,...,r_8$ are the
eight eighth roots of $-1$. The wavelengths $\lambda_n$ and decay
lengths $\nu_n$ of the eight fundamental complex solutions
$\exp(r_n z/\ell)$ (with $r_n=2\pi i/\lambda_n+1/\nu_n$) are
comprised of the four possible combinations of
$\lambda_n=\pm16.419$ and $\nu_n=\pm1.0824$, and the four possible
combinations of $\lambda_n=\pm6.8009$ and $\nu_n=\pm2.6131$.

\begin{figure}
\includegraphics[height=2.7in]{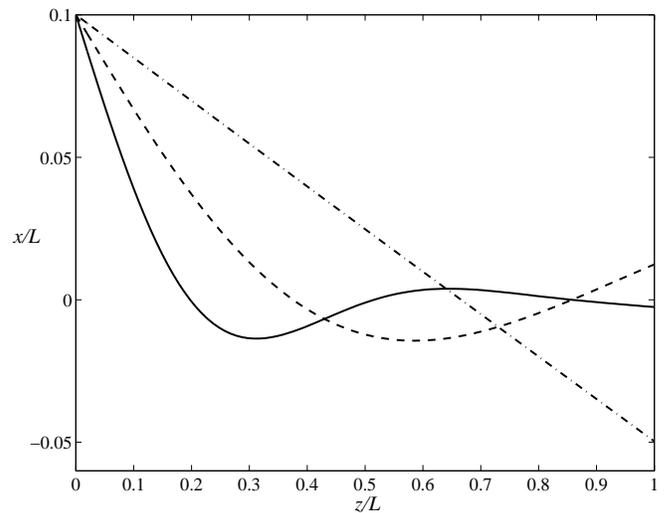}
\caption{\label{shapeszx} Shapes of a rotating flexible rod,
projected onto the $z$-$x$ plane, for $\ell/L=0.1$ (solid line),
$\ell/L=0.2$ (dashed line), and $\ell/L=0.5$ (dot-dashed line).
Vertical amplitudes have been exaggerated for clarity.}
\end{figure}

\begin{figure}
\includegraphics[height=2.7in]{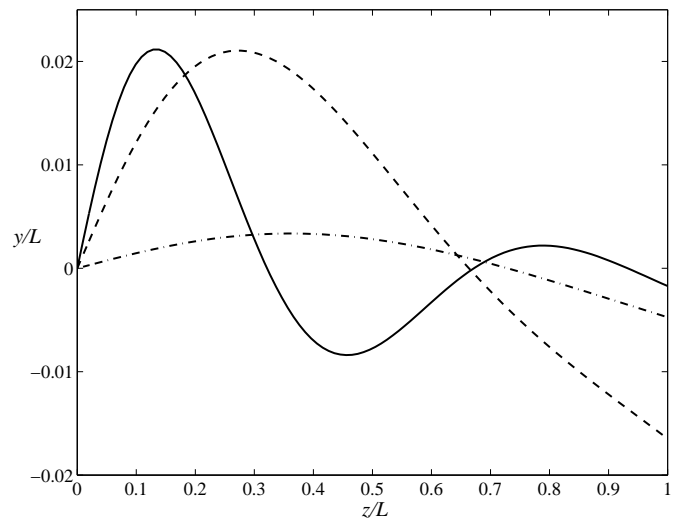}
\caption{\label{shapesyz} Shapes of a rotating flexible rod,
projected onto the $y$-$z$ plane, for $\ell/L=0.1$ (solid line),
$\ell/L=0.2$ (dashed line), and $\ell/L=0.5$ (dot-dashed line).
Vertical amplitudes have been exaggerated for clarity.}
\end{figure}

The boundary conditions determine the amplitudes and phases of the
coefficients $A_n$. At the distal end $z=L$, there is zero force
and moment: $A\partial^3\bm{r}_\perp/\partial z^3=0$,
$A\partial^2\bm{r}_\perp/\partial
z^2=0$~\cite{landau_lifshitz_elas}. At the proximal end, flagellar
filaments are connected to the rotary motor by a hook which is
more flexible than the rest of the filament.  We simply model this
flexible connection as a hinge with zero moment at $z=0$:
$A\partial^2\bm{r}_\perp/\partial z^2=0$.  (The other extreme, a
rigid hook with $\partial\bm{r}_\perp/\partial z=0$ leads to
qualitatively similar shapes for $\ell/L<1$, except near $z=0$.)
Finally, $\bm{r}_\perp(z=0)=b\hat{\bm{x}}$.  Applying these
boundary conditions to the solutions in eqn.~(\ref{machin_solns})
with $b=L/10$ yields the shapes shown in
figs.~\ref{shapesxy}--\ref{shapesyz}. For large $\ell$, the rod is
very stiff and does not bend; it is easy to show that in the limit
of $\ell/L\gg1$ that $x(z)=b(1-3z/2) +{\cal O}((L/\ell)^{4})$ and
$y(z)={\cal O}((L/\ell)^{4})$ for the hinged (zero moment)
boundary condition at $z=0$.  In the lab frame, the rod pivots
about the point $z/L=2/3$, tracing out a cone (a rigid rod
confined to the plane also pivots about this point in a viscous
fluid~\cite{wiggins_goldstein1998}).  When $\ell/L<1$, the rod
spirals around the axis of rotation, with the spiral becoming more
complete as $\ell$ gets smaller and smaller. Note the anisotropy;
the projection of the shape $x$-$y$ plane is elongated along the
$x$-axis (fig.~\ref{shapesxy}).  The rod configuration is a
compromise between minimizing the bending energy and minimizing
the dissipation rate. For $\ell/L\gg1$, elasticity dominates, and
the rod is straight. For $\ell/L\ll1$, viscous effects dominate
and the rod bends to align along the axis of rotation and thus
minimize the dissipation rate.  In this limit, the linear
approximation for the shape of the rod becomes invalid, and we
must replace (\ref{elastic_energy}) with the full expression for
the curvature. This nonlinear problem is readily solved by
standard methods (see {\it e.g.}~\cite{koehler_powers2000}); the
result is that inaccuracies of only a few percent arise when
$\ell/L\approx1/10$.

To assess the importance of the role of body rotation in bundling,
we estimate the characteristic length $\ell$. Various estimates
have appeared for the flagellar filament stiffness $A$, from
$10^{-24}$ N~m$^2$~\cite{fushime_maruyama_asakura1972} to
$10^{-22}$ N~m$^2$~\cite{hoshikawa_kamiya1985}.  Fortunately, the
characteristic length $\ell$ is not very sensitive to the value of
$A$. To estimate the transverse drag coefficient $\zeta_\perp$,
eqn.~(\ref{zetaperp}), we use the viscosity of water $\eta=0.001$
N~s/m$^2$, a typical length $L=10$ $\mu$m, and a diameter $2a=20$
nm. With a typical body rotation rate of $\omega_{\textrm{B}}=10$
Hz~\cite{macnab1977} and the range of stiffnesses quoted above,
the characteristic length $\ell$ is found to be two to six
microns. Therefore, the filaments are sufficiently flexible for
the observed body rotation rate to contribute significantly to
bundling.  Furthermore, the linear treatment of the rod shape is
justified. Presumably, body rotation is especially important for
the bundles which include many right-handed filaments and a single
left-handed filament, as observed in
ref.~\cite{turner_ryu_berg2000}.

Including axial drag does not alter the conclusions.  Axial drag
due to the swimming velocity leads to a tension gradient in the
rod which slightly increases the spiral pitch. Assuming a constant
tension equal to the maximum tension at the base of the rod and
disregarding the shadow effect of the cell body yields an upper
bound on the change in pitch. For a swimming velocity of about
$30$ $\mu$m/sec, the change in pitch is small compared to the
pitch.

The purpose of this work has been to point out the importance of
body rotation for flagellar filament bundling.  In order to focus
on the essential physics of this element of the bundling
phenomenon, we have disregarded several important but complementary
effects, such as
the helical shape of the flagellar filament and the flows induced
by the individual filaments~\cite{promise}. Despite these simplifications, 
we have shown that bacterial flagellar filaments are flexible
enough for body rotation to lead to wrapping.

\begin{acknowledgments}
I am indebted to R.E. Goldstein and G. Huber for important
conversations and ongoing collaborations.  This work is supported
by NSF grant no. CMS-0093658.
\end{acknowledgments}


\begin{thebibliography}{000}
\bibitem{macnab_ornston1977}
R.M. MacNab and M.K. Ornston, 
J. Mol. Biol. \textbf{112} (1977) 1.

\bibitem{bray1992}
D. Bray,  \textit{Cell Movements} (Garland Publishing, Inc.,
New York 1992).

\bibitem{berg1993}
H.C. Berg, \textit{Random Walks in Biology} (Princeton
University Press, Princeton, 1993).

\bibitem{turner_ryu_berg2000}
L. Turner, W.S. Ryu, and H.C. Berg,  J. Bacteriol.
\textbf{182} (2000) 2793. 

\bibitem{anderson1975}
R.A. Anderson, in  \textit{Swimming and Flying in Nature,}
vol.~\textbf{1}, edited by T.Y.-T. Wu, C.J. Brokaw, and C. Brenner
(Plenum Press, New York, 1975).

\bibitem{macnab1977}
R.M. MacNab, Proc. Natl. Acad. Sci. USA
\textbf{74} (1977) 221. 

\bibitem{lighthill1975}
J. Lighthill, J. SIAM Review \textbf{18} (1975)
161. 

\bibitem{promise}
A.J. Van Parys, K.S. Breuer,  and T.R. Powers, unpublished.

\bibitem{machin1958}
K.E. Machin, J. Exp. Biol. \textbf{35} (1958) 796. 

\bibitem{wiggins_goldstein1998}
C.H. Wiggins and R.E. Goldstein  Phys. Rev. Lett.
\textbf{80} (1998) 3879. 

\bibitem{happel_brenner1965}
J. Happel and H. Brenner, \textit{Low Reynolds Number
Hydrodynamics} (Prentice-Hall, Englewood Cliffs, NJ, 1965).



\bibitem{higdon}
J.J.L. Higdon, J. Fluid Mech. \textbf{90} (1979) 685. 

\bibitem{wolgemuth_et_al2000a}
C.W. Wolgemuth, T.R. Powers, and R.E. Goldstein, 
Phys. Rev. Lett. \textbf{84} (2000) 1623. 

\bibitem{landau_lifshitzFM}
L.D. Landau and E.M. Lifshitz, \textit{Fluid mechanics},
(Butterworth Heinemann, Oxford, 2nd edition, 1987).

\bibitem{landau_lifshitz_elas}
L.D. Landau and E.M. Lifshitz, \textit{Theory of
elasticity,} (Pergamon Press, Oxford, 3rd edition, 1986).

\bibitem{koehler_powers2000}
S.A. Koehler and T.R. Powers, Phys. Rev. Lett.
\textbf{85} (2000) 4827. 

\bibitem{fushime_maruyama_asakura1972}
S. Fujime, M. Maruyama, and S. Asakura,  J. Mol.
Biol. \textbf{68} (1972) 347. 

\bibitem{hoshikawa_kamiya1985}
H. Hoshikawa, and R. Kamiya, Biophys. Chem.
\textbf{22} (1985) 159. 

\end{thebibliography}


\end{document}